\documentclass{ifacconf}

\usepackage{amsmath,amsfonts,amssymb,graphicx,stmaryrd}
\usepackage{graphics}
\usepackage{natbib}        % required for bibliography

\def\ee{{(11)}}
\def\et{{(12)}}
\def\te{{(21)}}
\def\tt{{(22)}}

\newcommand{\bd}{\begin{displaymath}}
\newcommand{\ed}{\end{displaymath}}
\newcommand{\bea}{\begin{eqnarray}}
\newcommand{\eea}{\end{eqnarray}}
\newcommand{\Hi}{{\cal H}_\infty}

\newcommand{\R}{\mathbb{R}}
\newcommand{\C}{\mathbb{C}}

\newcommand{\w}{\omega}

\newcommand{\UU}{\mathbf{U}}
\newcommand{\VV}{\mathbf{V}}

\gdef \RR{{\Bbb R}}

\gdef \ZZ{{\Bbb Z}}

\makeatletter
\def\Ddots{\mathinner{\mkern1mu\raise\p@
\vbox{\kern7\p@\hbox{.}}\mkern2mu
\raise4\p@\hbox{.}\mkern2mu\raise7\p@\hbox{.}\mkern1mu}}
\makeatother

\begin{document}
\begin{frontmatter}

\title{On the sensitivity of the $\Hi$ norm of systems described by delay differential algebraic equations}

\author[First]{Suat Gumussoy},
\author[First]{Wim Michiels}

\address[First]{Department of Computer Science, K. U. Leuven, \\
        Celestijnenlaan 200A, 3001, Heverlee, Belgium \\
        \mbox{(e-mails: suat.gumussoy@cs.kuleuven.be, wim.michiels@cs.kuleuven.be)}.}

\begin{abstract}
We consider delay differential algebraic equations (DDAEs) to model interconnected systems with time-delays. The DDAE framework does not require any elimination techniques and can directly deal with any interconnection of systems and controllers with time-delays. In this framework, we analyze the properties of the $\Hi$ norm of systems described by delay differential algebraic equations. We show that the standard $\Hi$ norm may be sensitive to arbitrarily small delay perturbations. We introduce the strong $\Hi$ norm which is insensitive to small delay perturbations and describe its properties. We conclude that the strong $\Hi$ norm is more appropriate in any practical control application compared to the standard $\Hi$ norm for systems with time-delays whenever there are high-frequency paths in control loops.
\end{abstract}


\end{frontmatter}
%===============================================================================

\section{Introduction}
In robust control applications, the design requirements are usually defined in terms of $\Hi$ norms of the closed-loop functions including the plant, the controller and weights for uncertainties and disturbances \cite{zhou}. The properties and robust computational methods of the $\Hi$ norm of closed-loop functions are  essential in a computer aided control system design. The properties of $\Hi$ norm for finite dimensional multi-input-multi-output systems are well-known and reliable numerical methods for the $\Hi$ norm computation are available \cite{boydbala,steinbuch}.

We analyze the sensitivity of the $\Hi$ norm of systems described by delay differential algebraic equations.  An important motivation for systems under consideration stems from the fact that interconnected systems with delays can be naturally modeled by state-space representation of the form
\begin{equation}\label{system}
\left\{\begin{array}{l}
E \dot x(t)= A_0 x(t)+\sum_{i=1}^m A_i x(t-\tau_i) +B w(t), \\
z=C x(t).
\end{array}\right.
\end{equation}
The time-delays $\tau_i$, $i=1,\ldots,m$ are positive real numbers. The system matrices are $E$ and $A_i$, $i=0,\ldots,m$ are real-valued square matrices and other system matrices with the capital letters are real-valued matrices with appropriate dimensions. The input $w$ and output $z$ are disturbances and signals to be minimized to achieve design requirements and some of system matrices may include the controller parameters.

The system with the closed-loop equations (\ref{system}) represents all interesting cases of the feedback connection of a time-delay plant and a controller. The transformation of the closed-loop system to this form can be easily done by first augmenting the system equations of the plant and controller. As we shall see, this augmented system can subsequently be brought in the form (\ref{system}) by introducing slack variables to eliminate input/output delays and direct feedthrough terms in the closed-loop equations. Hence, the resulting system of the form (\ref{system}) is obtained directly without complicated elimination techniques, that may even not be possible in the presence of time-delays. It can serve as a standard form for the development of control design and software.

%Our approach for the fixed-order $\Hi$ controller design is based on a non-smooth, non-convex optimization method \cite{suatHIFOO} and tunes the controller parameters in the system matrices of (\ref{system}) in order minimize the $\Hi$ norm of (\ref{system}). This approach requires the computation of $\Hi$ norms  for DDAEs and their derivatives with respect to the controller parameters.

By interconnecting systems and controller high frequency paths could be created in control loops, which may lead to sensitivity problems with respect to the delays and delay perturbations. Therefore it is important to take the sensitivity explicitly into account in the design. We will illustrate that the $\Hi$ norm of the transfer function from $w$ to $z$ in (\ref{system}) may be sensitive to arbitrarily small delay changes. Since small modeling errors are inevitable in any practical design we are interested in the smallest upper bound of the $\Hi$ norm that is insensitive to small delay changes. Inspired by the concept of strong stability of neutral equations \cite{have:02}, this leads us to the introduction of the concept of  \emph{strong $\Hi$ norms} for DDAEs,  Several properties of the strong $\Hi$ norm are shown and a computational formula is obtained. The theory derived can be considered as the dual of the theory of strong stability as elaborated in \cite{have:02,TW-report-286,Michiels:2005:NEUTRAL,Michiels:2007:MULTIVARIATE} and the references therein.

The characterization of the $\Hi$ norm is frequency domain based and builds on the eigenvalue based framework developed in~\cite{bookwim}.  Time-domain approach for the $\Hi$ control of DDAEs have been described in \cite{fridman} and the references therein, based on the construction of Lyapunov-Krasovskii functionals.

\smallskip

The structure of the article is as follows. In Section \ref{sec:motex} we illustrate the generality of the system description (\ref{system}). Preliminaries and assumptions are given in Section \ref{sec:prelim}. The definition and properties of the strong $\Hi$ norm of DDAE are given in Section \ref{sec:shinf}. Section \ref{sec:ex} is devoted to the numerical examples. In Section \ref{sec:conc} some concluding remarks are presented.

\subsection*{Notations} The notations are
as follows:
\begin{tabbing}
  \= $j$\hspace{1.1cm} \=:  the imaginary identity \\
  \> $\vec\tau$ \>:  vector notation for $(\tau_1,\ldots,\tau_m)$ \\
  \> $\C, \R, \mathbb{N}$  \>: set of the complex, real and natural numbers \\
  \> $\R^+,\R_0^+$ \> : set of  nonnegative and strictly positive real \\
  \> \>  \; numbers \\
  \> $A^{-T}$ \>: transpose of the inverse matrix of $A$ \\
  \> $A^{\perp}$ \>: matrix of full column rank whose columns  \\
  \> \> \ \  span the
   orthogonal complement of the    \\
  \> \> \ \ nullspace of $A$ \\
  \> $0, I$ \> : zero and identity matrices of appropriate\\
  \>  \>  \; dimensions\\
  \> $\sigma_i(A)$ \>: i$^\mathrm{th}$ singular value of $A$,\ $\sigma_1(\cdot)\geq\sigma_2(\cdot)\geq \cdots$ \\
  \> $\Re(u)$ \> : real part of the complex number $u$ \\
  \> $\mathcal{B}(\vec \tau,\epsilon)$ \> : open ball of radius $\epsilon\in\R^+$ centered at $\vec\tau\in$ \\
  \> \> \ \ $(\R^+)^m$, $\mathcal{B}(\vec \tau,\epsilon):=\{\vec\theta\in(\R)^m : \|\vec\theta-\vec \tau\|<\epsilon\}$
\end{tabbing}

\section{Motivating examples} \label{sec:motex}

With some simple examples we illustrate the generality of the system description (\ref{system}).

\begin{exmp} \label{elim:connect}
Consider the feedback interconnection of the system
\[
\left\{\begin{array}{lll}
\dot x(t)&=&A x(t)+B_1 u(t)+B_2w(t),\\
 y(t)&=& C x(t)+D_1 u(t),\\
 z(t)&=& F x(t),
\end{array}\right.
\]
and the controller
\[
u(t)=K y(t-\tau).
\]
For $\tau=0$ it is possible to eliminate the output and controller equation, which results in the closed-loop system
\begin{equation}\label{elimination}
\left\{\begin{array}{lll}
\dot x(t)&=& A x(t)+B_1 K (I-D_1 K)^{-1} C x(t)+B_2 w(t), \\
z(t) & =& F x(t).
\end{array}\right.
\end{equation}
This approach is for instance taken in the software package HIFOO~\cite{Burke-hifoo}.
If $\tau\neq 0$, then the elimination is not possible any more. However, if we let $X=[x^T\ u^T y^T]^T$ we can describe the system by the equations
{\small\[
\left\{\begin{array}{l}
\left[\begin{array}{ccc}
I & 0 & 0 \\ 0 &0 & 0 \\ 0& 0&0
\end{array}\right]\dot X(t)=
\left[\begin{array}{ccc}
A & B_1 & 0 \\
C & D_1 &-I\\
0& I & 0
\end{array}\right] X(t)+
\left[\begin{array}{ccc}
0 & 0 & 0 \\
0 & 0 & 0\\
0& 0 & -K
\end{array}\right] X(t-\tau) \\
\hspace{6.5cm} +\left[\begin{array}{c}B_2\\0\\0
\end{array}\right] w(t),
\\
z(t)=\left[\begin{array}{cc c} F & 0&0  \end{array}\right]X(t),
 \end{array}\right.
 \]} which are of the form (\ref{system}).  Furthermore, the dependence of the matrices of the closed-loop system on the controller parameters, $K$, is still linear, unlike in (\ref{elimination}).
\end{exmp}
\begin{exmp} \label{elim:feedthru}
The presence of a direct feedthrough term from $w$ to $z$, as in
\begin{equation}\label{ex2}
\left\{\begin{array}{lll}
\dot x(t)&=& Ax(t)+A_1 x(t-\tau)+B w(t),\\
z(t)&=&F x(t)+D_2 w(t),
\end{array}\right.
\end{equation}
can be avoided by introducing a slack variable. If we let $X=[x^T\ \gamma_w^T]^T$, where $\gamma_w$ is the slack variable, we can bring (\ref{ex2}) in the form (\ref{system}):
{\small \[
\left\{\begin{array}{l}
\left[\begin{array}{cc}
I & 0 \\ 0 &0
\end{array}\right]\dot X(t)=
\left[\begin{array}{cc}
A & 0 \\ 0 & -I
\end{array}\right] X(t) +
\left[\begin{array}{cc}
 A_1 & 0 \\ 0 & 0
\end{array}\right] X(t-\tau)+
\left[ \begin{array}{l} B\\ I
\end{array}\right] w(t),
\\
z(t)=[F\ D_2]\ X(t).
\end{array}\right.
\]}
\end{exmp}
\begin{exmp} \label{elim:inputdelay}
The system
\[
\left\{\begin{array}{lll}
\dot x(t) &=& A x(t)+B_1 w(t)+B_2 w(t-\tau),\\
z(t)&=& C x(t),
\end{array}\right.
\]
can also be brought in the standard form (\ref{system}) by a slack variable. Letting $X=[x^T \gamma_w^T]^T$ we can express
{\small \[
\left\{\begin{array}{lll}
\dot X(t) &=&
\left[\begin{array}{cc}
A & B_1 \\
0 & -I
\end{array}\right]
X(t)
+
\left[\begin{array}{cc}
0 & B_2 \\
0 & 0
\end{array}\right]
X(t-\tau)
+
\left[\begin{array}{c}
0 \\ I
\end{array}\right] w(t),
\\
z(t)&=& [C\ \  0]\ X(t).
\end{array}\right.
\]}
In a similar way one can deal with delays in the output~$z$.
\end{exmp}

Using the techniques illustrated with the above examples a broad class of interconnected
systems with delays can be brought in the form (\ref{system}), where the external
inputs $w$ and outputs $z$ stem from the performance specifications expressed in terms of
appropriately defined transfer functions. The price to pay for the generality of the framework is the increase of the dimension of the system, $n$, which affects the efficiency of the numerical methods. However, this is a minor problem in most applications because the delay difference equations or algebraic constraints are related to inputs and outputs, and
the number of inputs and outputs is usually much smaller than the number of state variables.

Finally, we note that also neutral time-delay systems can be directly dealt with,  as shown in the following example.
\begin{exmp} \label{elim:inputdelay}
The neutral time-delay system
\[
\left\{\begin{array}{rll}
\frac{d}{dt}\left(x(t)+Dx(t-\tau_1)\right) &=& A_0 x(t)+A_1 x(t-\tau_2)+B w(t),\\
z(t)&=& C x(t),
\end{array}\right.
\]
can be represented in the form (\ref{system}) using a slack variable. If we let $X=[x^T \gamma_x^T]^T$ we can describe the system by the equations
{\small \[
\left\{\begin{array}{rll}
\left[\begin{array}{cc}
0 & I \\
0 & 0
\end{array}\right] \dot X(t) &=&
\left[\begin{array}{cc}
A_0 & 0 \\
I & -I
\end{array}\right]
X(t)
+
\left[\begin{array}{cc}
0 & 0 \\
D & 0
\end{array}\right]
X(t-\tau_1)
+ \\
& & \hspace{1cm}\left[\begin{array}{cc}
A_1 & 0 \\
0 & 0
\end{array}\right]
X(t-\tau_2)
+
\left[\begin{array}{c}
B \\ 0
\end{array}\right] w(t),
\\
z(t)&=& [C\ \  0]\ X(t).
\end{array}\right.
\]}
\end{exmp}

\section{Definitions and Assumptions} \label{sec:prelim}
\subsection*{Assumptions}
Let $\mathrm{rank}(E)=n-m$, with $m\leq n$, and let the columns
of matrix $U\in\RR^{n\times m}$, respectively $V\in\RR^{n\times m}$, be a (minimal) basis for
the left, respectively right nullspace, that is,
\begin{equation}\label{nullspace}
U^T E=0,\ \ E V=0.
\end{equation}
Throughout the paper we make the following assumption.
\begin{assum} \label{assumption}
The matrix $U^T A_0 V$ is nonsingular.
\end{assum}

\medskip

In order to motivate Assumption \ref{assumption}, we note that the equations (\ref{system}) can be separated into coupled
delay differential and delay  difference equations. When we
define
\[
\mathbf{U}= \left[{U^{\perp}}\  U\right],\ \ \mathbf{V}=\left[V^{\perp}\ V\right],\
\]
a pre-multiplication of (\ref{system}) with $\UU^T$ and the substitution
\[
x=\VV\ [x_1^T\ x_2^T]^T,
\]
with $x_1(t)\in\RR^{n-m}$ and $x_2(t)\in\RR^m$, yield the coupled equations
{\small \begin{equation}\label{coupled}
\left\{\begin{array}{c}
%
%\begin{eqnarray}
E^\ee \dot x_1(t)= \sum_{i=0}^m A_i^\ee x_1(t-\tau_i) +\sum_{i=0}^m A_i^\et x_2(t-\tau_i)+B_1 w(t), \\
0=A_0^\tt x_2(t)+ \sum_{i=1}^m A_i^\tt x_2(t-\tau_i)+\sum_{i=0}^m A_i^\te x_1(t-\tau_i)+B_2 w(t), \\
y(t)=C_1 x_1(t)+C_2 x_2(t),
\end{array}\right.
\end{equation}}
%
%
%\end{eqnarray}
where
\begin{equation}\label{transmat}
\left.
\begin{array}{lll}
A_i^\ee= {U^{\perp}}^T A_i V^{\perp}, & A_i^\et= {U^{\perp}}^T A_i V,&\\
A_i^\te= {U}^T A_i V^{\perp}, & A_i^\tt= {U}^T A_i V, & i=0,\ldots,m,
\end{array}
\right.
\end{equation}
and
\begin{multline}\label{transmat2}
E^\ee= {U^{\perp}}^T E V^{\perp},\ \   B_1={U^{\perp}}^T B,\ \ B_2= U^T B, \\
C_1=C V^{\perp},\ \ C_2=C V.
\end{multline}
Matrix $E^\ee$ in (\ref{coupled}) is invertible, following from
\[
n-m=\mathrm{rank}(E)=\mathrm{rank}(\mathbf{U}^T E\mathbf{V})=\mathrm{rank}(E^{(11)}).
\]
In addition, matrix $A_0^{(22)}$ is invertible, following from
Assumption~\ref{assumption}.

The equations (\ref{coupled}) with $w\equiv 0$ are semi-explicit delay differential algebraic equations of index~1, because  delay differential equations are obtained by differentiating the second equation. This precludes the occurrence of impulsive solutions~\cite{fridman}. Moreover, the invertibility of $A_0^{(22)}$
 prevents that the equations are of \emph{advanced} type and, hence, non-causal. This further motivates
why Assumption~\ref{assumption} is natural in the delay case considered, although it restricts the index to
one (for a general treatment in the delay free case, see for instance~\cite{stykel} and the references therein).
\medskip

We also make the following assumption.
\begin{assum} \label{assumption_sstab}
The zero solution of system (\ref{system}), with $w\equiv0$, is strongly exponentially stable.
\end{assum}

Strong exponential stability refers to the fact that the asymptotic stability of the null solution is robust against small delay perturbations \cite{have:02,Michiels:2007:MULTIVARIATE}. Due to the modeling errors and uncertainty, the delays of the time-delay model are typically not exactly known and this type of stability is required in practice. The stability of the closed-loop system (\ref{system}) is a necessary assumption since the $\Hi$ norm is defined for stable systems only.

\subsection*{Transfer functions}

From (\ref{coupled}) we can write the transfer function of the system (\ref{system}) as
\begin{eqnarray}\label{transferT}
\label{T} \hspace{-.5cm} T(\lambda)&:=&C\left(\lambda E-A_0-\sum_{i=1}^m A_i e^{-\lambda \tau_i}\right)^{-1}B, \\
&=&[C_1 \ \ C_2]\left[\begin{array}{rr}\lambda E^\ee-A_{11}(\lambda) & -A_{12}(\lambda)\\
-A_{21}(\lambda) & -A_{22}(\lambda)
 \end{array}\right]^{-1} \left[\begin{array}{c}B_1\\ B_2 \end{array}\right], \label{transferblock}
\end{eqnarray}
with
\[
A_{kl}(\lambda)=\sum_{i=0}^m A_i^{(kl)} e^{-\lambda\tau_i},\ \ k,l\in\{1,2\}.
\]

We define the {\it asymptotic} transfer function of the system (\ref{system}) as
\begin{eqnarray}
\label{Ta} \hspace{-.6cm}T_a(\lambda) &:=&-C V \left(U^T A_0 V +\sum_{i=1}^m U^T A_i V e^{-\lambda\tau_i} \right)^{-1} U^TB\\
             &=&-C_2 A_{22}(\lambda)^{-1} B_2 .
\end{eqnarray}

The terminology stems from the fact that the transfer function $T$ and the asymptotic transfer function $T_a$ converge to each other for high frequencies, as precisized in the following Proposition.
\begin{prop}\label{propconverge}
$\forall\gamma >0$, $\exists\Omega>0$: $\sigma_{1}\left(T(j\w)-T_a(j\w)\right)<\gamma,\ \forall\w>\Omega$.
\end{prop}
\noindent\textbf{Proof.\ } The assertion follows from the explicit expression for the inverse of the two-by-two block matrix in (\ref{transferblock}), combined with the property that
 \begin{equation}\label{normA22}
\sup_{\Re(\lambda)\geq 0}\left \| \left( A_{22}(\lambda) \right)^{-1} \right\|_2
 \end{equation}
 if finite. The latter is due to Assumption~\ref{assumption_sstab}. \hfill $\Box$

The $\Hi$ norm of the transfer function $T$ of the \emph{stable} system (\ref{system}), is defined as
\begin{equation} \label{hinfnorm}
\|T(j\w)\|_\infty:=\sup_{\w\in j\R} \sigma_{1} \left( T(j\w) \right).
\end{equation}
Similarly we can define $\Hi$ norm of $T_a$.
\smallskip

\section{Strong $\Hi$ Norm of Time-Delay Systems} \label{sec:shinf}

We now analyze continuity properties of the $\Hi$ norm of the transfer function $T$ with respect to the delay parameters. The function
\begin{equation}\label{defTdel}
\vec\tau\in(\RR_{0}^+)^{m}\mapsto \|T(j\w,\vec\tau)\|_\infty
\end{equation}
is, in general, not continuous, which is inherited from the behavior of the asymptotic transfer function, $T_a$, more precisely the function
\begin{equation}\label{defTadel}
\vec\tau\in(\RR_{0}^+)^{m}\mapsto \|T_a(j\w,\vec\tau)\|_\infty.
\end{equation}
We start with a motivating example

\begin{figure}[!h]
 \begin{minipage}[t]{0.24\textwidth}
  \vspace{0pt}
   \includegraphics[width=\linewidth]{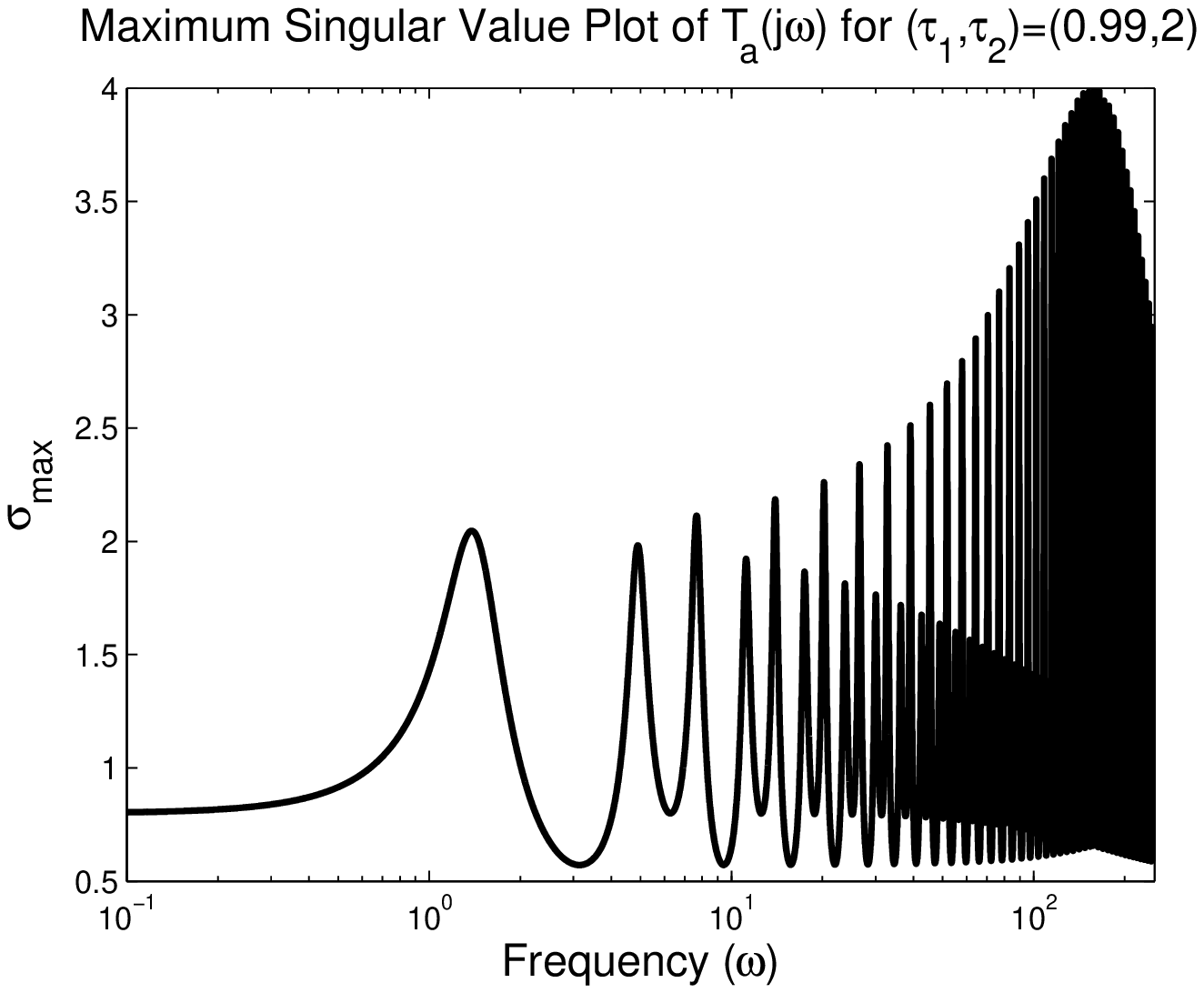}
 \end{minipage}
\hfill
 \begin{minipage}[t]{0.24\textwidth}
  \vspace{0pt}\raggedright
   \includegraphics[width=\linewidth]{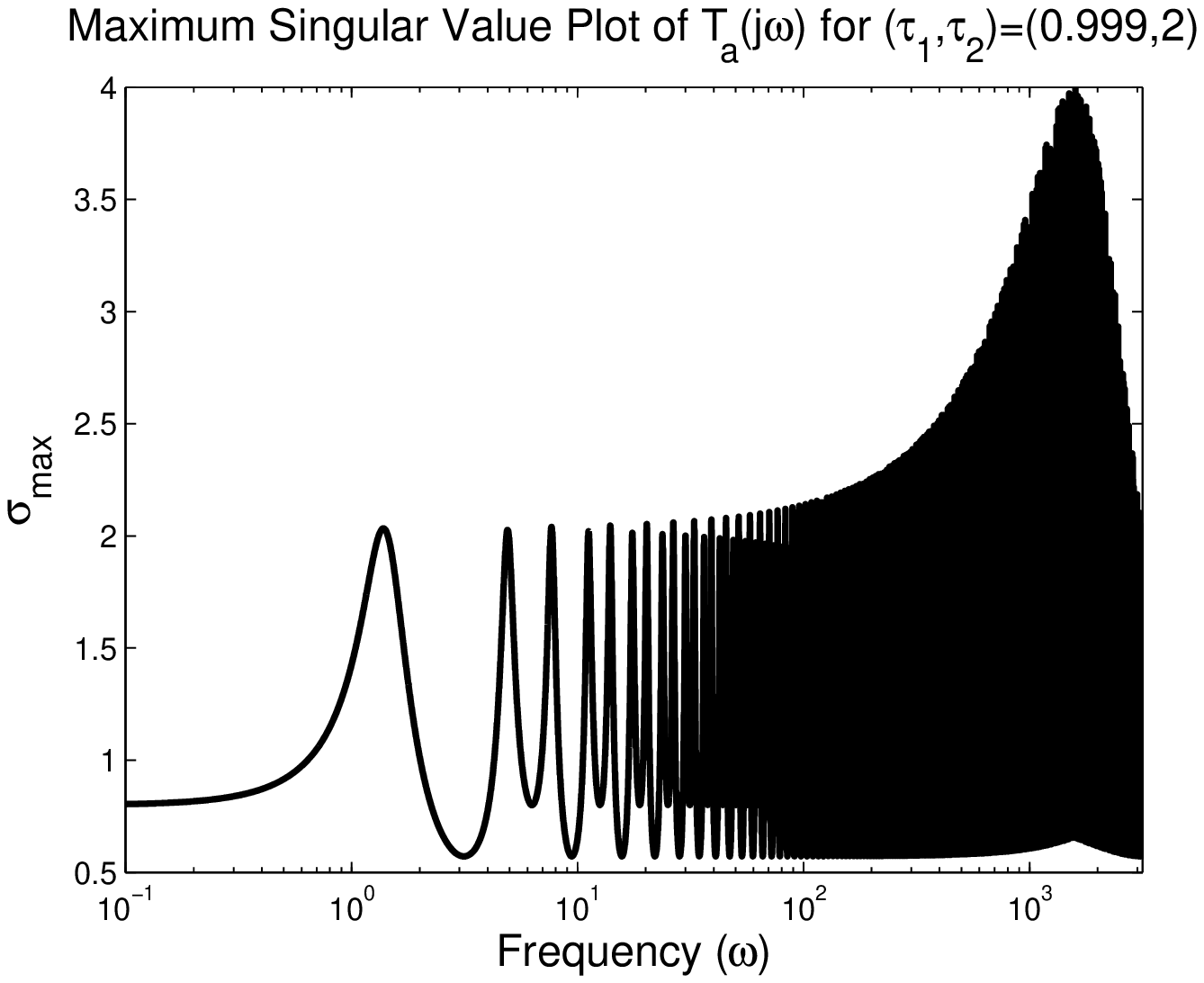}
 \end{minipage}
       \caption{\label{fig:svd2} The maximum singular value plot of $T_a(j\w)$ for $(\tau_1,\tau_2)=(0.99,2)$ (left) and $(\tau_1,\tau_2)=(0.999,2)$ (right) as a function of $\omega$.}
\end{figure}

\begin{exmp} \label{ex:TandTa}
Let the transfer function $T$ be defined as
\begin{equation} \label{Tex}
T(\lambda)=\frac{\lambda+2}{\lambda(1-0.25e^{-\lambda\tau_1}+0.5e^{-\lambda\tau_2})+1}
\end{equation} where $(\tau_1,\tau_2)=(1,2)$. The transfer function $T$ is stable, its $\Hi$ norm is $2.6422$ achieved at $\w=1.6598$ and the maximum singular value plot is given in Figure \ref{fig:svd1} (on the left). The high frequency behavior is described by the asymptotic transfer function
\begin{equation} \label{Taex}
T_a(\lambda)=\frac{1}{(1-0.25e^{-\lambda\tau_1}+0.5e^{-\lambda\tau_2})},
\end{equation}
whose $\Hi$ norm is equal to $2.0320$, which is less than $\|T(j\w)\|_\infty$. However, when the first time delay is perturbed to $\tau_1=0.99$, the $\Hi$ norm of the transfer function $T$ is $3.9993$, reached at $\w=158.6578$, see Figure \ref{fig:svd1} (on the right). The $\Hi$ norm of $T$ is quite different from that for $(\tau_1,\tau_2)=(1,2)$. A closer look at the maximum singular value plot of the asymptotic transfer function $T_a$ in Figure \ref{fig:svd2} (on the left) shows that the sensitivity is due to the transfer function $T_a$.

\begin{figure}[!h]
 \begin{minipage}[t]{0.24\textwidth}
  \vspace{0pt}
   \includegraphics[width=\linewidth]{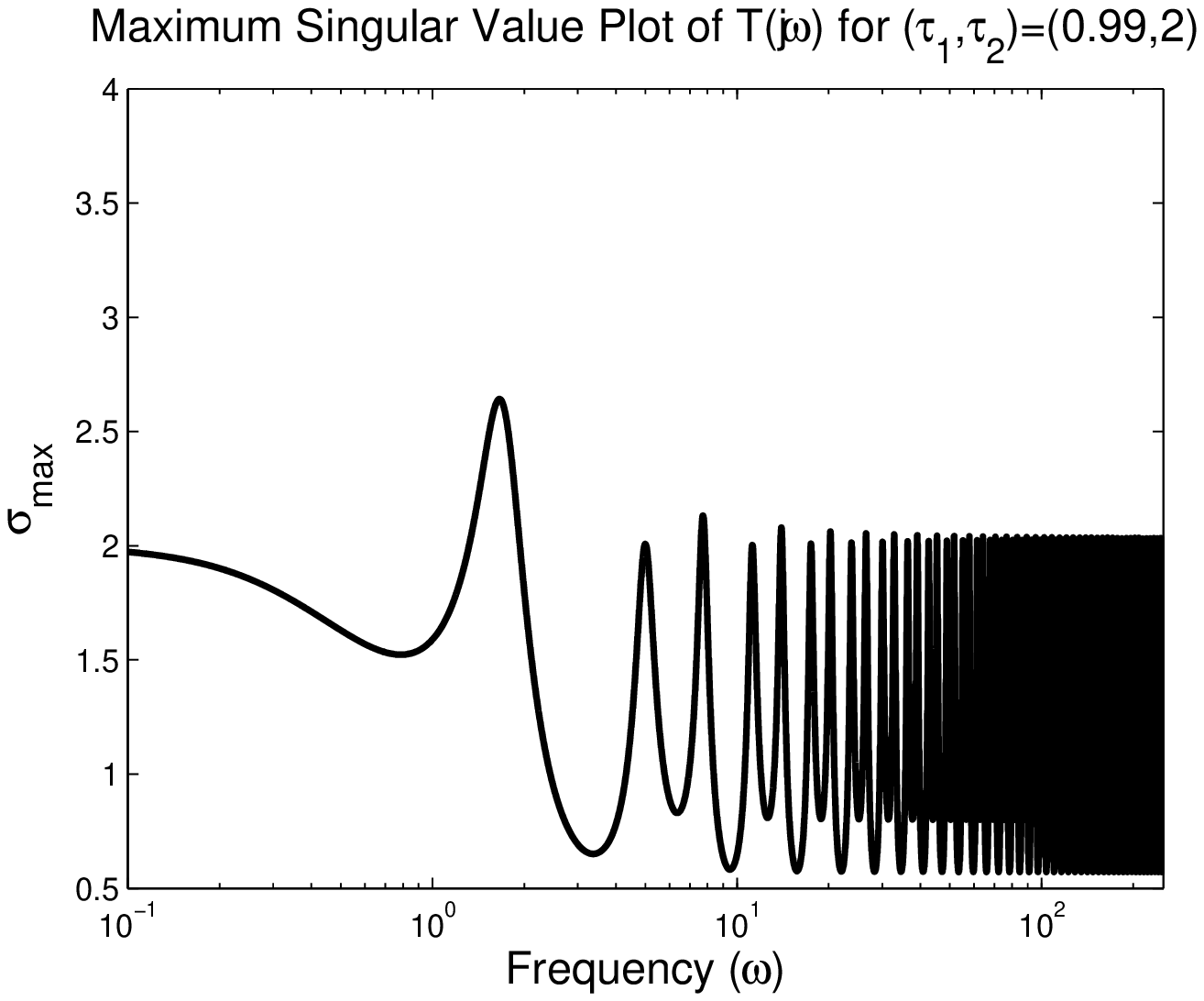}
 \end{minipage}
\hfill
 \begin{minipage}[t]{0.24\textwidth}
  \vspace{0pt}\raggedright
   \includegraphics[width=\linewidth]{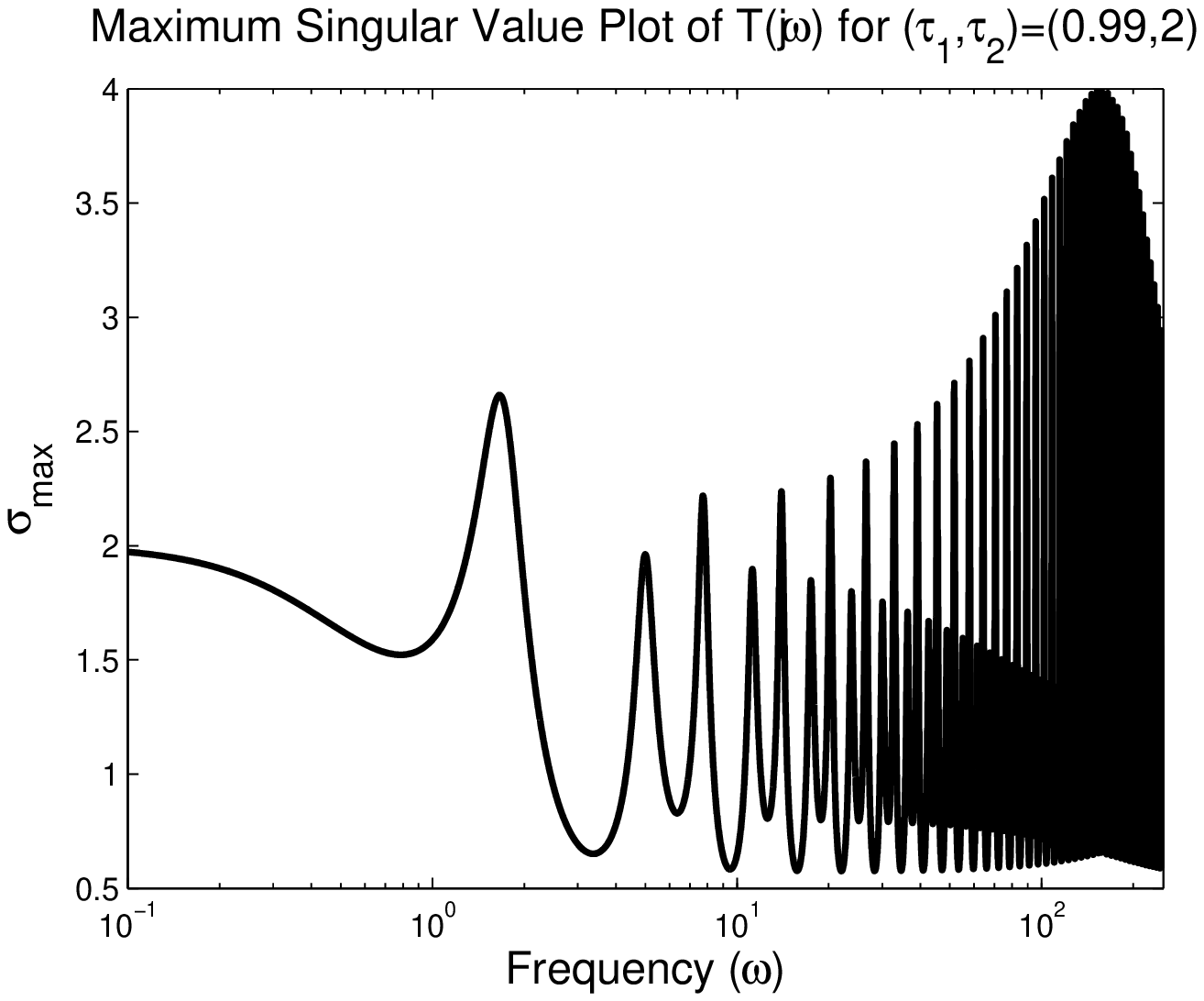}
 \end{minipage}
       \caption{\label{fig:svd1} The maximum singular value plot of $T(j\w)$ for $(\tau_1,\tau_2)=(1,2)$ (left) and $(\tau_1,\tau_2)=(0.99,2)$ (right) as a function of $\omega$.}
\end{figure}

Even if the first delay is perturbed slightly to $\tau_1=0.999$, the problem is not resolved, indicating that the functions (\ref{defTdel}) and (\ref{defTadel}) are discontinuous at $(\tau_1,\tau_2)=(1,2)$. The $\Hi$ norm of the transfer function $T$  for $(\tau_1,\tau_2)=(0.999,2)$ is namely given by $3.9998$, and the peak value is reached at  $\w=1566.0816$. The corresponding asymptotic transfer function $T_a$ is shown in Figure \ref{fig:svd2} (on the right). When the delay perturbation tends to zero, the frequency where the maximum in the singular value plot of the asymptotic transfer function $T_a$ is achieved moves towards infinity.
\end{exmp}

The above example illustrates that the $\Hi$ norm of the transfer function $T$ may be sensitive to \emph{infinitesimal} delay changes. Since this property is related to the behavior of the transfer function at high frequencies and, hence, the asymptotic transfer function $T_a$, we first study the properties of the function (\ref{defTadel}).

Since small modeling errors and uncertainty are inevitable in a practical design, we wish to characterize the smallest upper bound for the $\Hi$ norm of the asymptotic transfer function $T_a$ which is {\it insensitive} to small delay changes.

\begin{defn} \label{def:shinfTa}
For $\vec \tau\in(\RR_{0}^+)^{m}$, let the strong $\mathcal{H}_{\infty}$ norm of $T_a$, $\interleave {T_a}(j\w,\vec \tau)\interleave_\infty$, be defined as
\begin{multline} \nonumber
\interleave T_a(j\w,\vec \tau)\interleave_\infty:=\lim_{\epsilon\rightarrow 0+}
\sup \{\|T_a(j\w,\vec \tau_\epsilon)\|_\infty: \\ \vec \tau_\epsilon\in\mathcal{B}(\vec \tau,\epsilon) \cap (\RR^+)^{m} \}.
\end{multline}
\end{defn}

Several properties of this upper bound on $\|T_a(j\w,\vec \tau)\|_\infty$ are listed below.
\begin{prop}\label{prop:Tasinfprop}
The following assertions hold:
\begin{enumerate}
\item for every $\vec\tau\in(\RR_0^+)^m$, we have
\begin{equation} \label{Tasweep}
\interleave T_a(j\w,\vec \tau)\interleave_\infty=\max_{\vec\theta\in [0,\
2\pi]^m} \sigma_{1} \left( \mathbb{T}_a(\vec \theta) \right),
\end{equation}
where
\begin{eqnarray} \label{Ta_theta}
\mathbb{T}_a(\vec \theta)&=&C_2 \left(-A_0^{(22)} -\sum_{i=1}^m A_i^{(22)} e^{-j\theta_i} \right)^{-1} B_2, \\
\nonumber &=&C V \left(-U^T A_0 V -\sum_{i=1}^m U^T A_i V e^{-j\theta_i} \right)^{-1} U^TB.
\end{eqnarray}
\item $\interleave T_a(j\w,\vec \tau)\interleave_\infty \geq \|T_a(j\w,\vec \tau)\|_{\infty}$ for all delays $\vec \tau$;
\item $\interleave T_a(j\w,\vec \tau)\interleave_\infty=\|T_a(j\w,\vec \tau)\|_\infty$ for rationally
independent\footnote{The $m$ components of
$\vec\tau=(\tau_1,\ldots,\tau_m)$ are rationally
independent if and only if $\sum_{k=1}^m z_k \tau_k=0,\
z_k\in\ZZ$ implies $z_k=0,\ \forall k=1,\ldots,m$. For
instance, two delays $\tau_1$ and $\tau_2$ are rationally
independent if their ratio is an irrational number.} $\vec \tau$.
\end{enumerate}
\end{prop}
\noindent\textbf{Proof.\ } We always have
\begin{multline}\nonumber
(e^{-j\w\tau_1},\ldots,e^{-j\w\tau_m})\in \{(e^{-j\theta_1},\ldots,e^{-j\theta_m}): \\
\theta_i\in[0,2\pi], \ i=1,\ldots,m  \},
\end{multline}
implying
\begin{equation}\label{arg1}
\|T(j\omega,\vec\tau)\|_{\infty}\leq \max_{\vec\theta\in [0,\
2\pi]^m} \sigma_{1} \left( \mathbb{T}_a(\vec \theta) \right).
\end{equation}
For any $\epsilon>0$ in Definition \ref{def:shinfTa}, there exists $\vec \tau_\epsilon=[ \tau_{\epsilon,1},\ldots,\tau_{\epsilon,m}]$ rationally independent in $\mathcal{B}(\vec \tau,\epsilon)\cap(\RR^+)^m$. By Theorem $2.1$ in \cite{TW-report-286}, given rationally independent time delays $\vec \tau_\epsilon$ and for $\vec \theta=[\theta_1,\ldots,\theta_m]$ arbitrary, there exists a sequence of real numbers $\{\w_n\}_{n\geq1}$ such that
\[
\lim_{n\rightarrow\infty} \max_{1\leq i \leq m} \left|e^{-j\w_n \tau_{\epsilon,i}}-e^{-j\theta_i}\right|=0.
\]
It follows that
\begin{multline}
\nonumber \mathrm{closure}\{(e^{-j\w\tau_{\epsilon,1}},\ldots,e^{-j\w\tau_{\epsilon,m}}): \w\in\R  \}= \\
\{(e^{-j\theta_1},\ldots,e^{-j\theta_m}): \theta_i\in[0,2\pi],  i=1,\ldots,m  \},
\end{multline}
implying
\begin{equation}\label{arg2}
\|T(j\omega,\vec\tau_{\epsilon})\|_{\infty}= \max_{\vec\theta\in [0,\
2\pi]^m} \sigma_{1} \left( \mathbb{T}_a(\vec \theta) \right).
\end{equation}
The assertions follow from (\ref{arg1}) and (\ref{arg2}).
%
%
%The first assertion (\ref{Tasweep}) follows. Since $\interleave T_a(j\w,\vec \tau)\interleave_\infty$ is computed over $\mathcal{B}(\vec \tau,\epsilon)$ including nominal delays $\vec \tau$ and does not depend on time-delay values, the second assertion follows. The third assertion is a direct consequence of the fact that for rationally independent $\vec \tau$, there exists a one-to-one mapping between $\w_0\vec\tau$ and $\vec \theta_0$ for all $\w_0\in\R$ and $\vec \theta_0\in[0,\ 2\pi]^m$. The third assertion follows.
\hfill $\Box$

Formula (\ref{Tasweep}) in Proposition \ref{prop:Tasinfprop} shows that the strong $\Hi$ norm is independent of the delay values. The formula further naturally leads to a computational scheme based on sweeping on $\vec \theta$ intervals. This approximation can be corrected by solving a set of nonlinear equations. Numerical computation details are presented in \cite{TW579:10}.

\medskip

We now come back to the properties of the transfer function (\ref{defTdel}) of the system (\ref{system}). As we have illustrated, a discontinuity of the function (\ref{defTadel}) may carry over to the function (\ref{defTdel}). Therefore, we define the strong $\Hi$ norm of the transfer function~$T$ in a similar way.
\begin{defn}
 For $\vec \tau\in(\RR_{0}^+)^{m}$,  the strong $\Hi$ norm of $T$, $\interleave {T}(j\w,\vec \tau)\interleave_\infty $, is given by
\begin{multline} \label{shinfnorm_def}
   \interleave T(j\w,\vec \tau)\interleave_\infty:=\lim_{\epsilon\rightarrow 0+}
\sup \{\|T(j\w,\vec \tau_\epsilon)\|_\infty: \\
\vec \tau_\epsilon\in\mathcal{B}(\vec \tau,\epsilon) \cap (\RR^+)^{m} \}.
\end{multline}
\end{defn}

The following main theorem describes, among others, the desirable property that, in contrast to the $\Hi$ norm, the strong $\Hi$ norm \emph{continuously} depends on the delay parameters. The proof makes use of the technical results in Section~\ref{sec:Appendix2} of the appendix.
\begin{thm}
%\begin{enumerate}
%\item
The strong $\Hi$ norm of the delay differential algebraic system (\ref{system}) satisfies
\begin{equation} \label{shinfnorm}
    \interleave T(j\w,\vec \tau)\interleave_\infty=\max\left(\|T(j\w,\vec\tau)\|_{\infty}, \interleave T_a(j\w,\vec \tau)\interleave_\infty \right),
\end{equation} where $T$ and $T_a$ are the transfer function (\ref{T}) and the asymptotic transfer function (\ref{Ta}).
%\item
%
%\item
In addition, the function
\begin{equation}\label{shinfnorm2}
    \vec \tau\in(\RR^+_0)^m\mapsto \interleave T(j\w,\vec \tau)\interleave_\infty
\end{equation}
is continuous.
%\end{enumerate}
\end{thm}

\noindent\textbf{Proof.\ }
Lemma~\ref{lem:fininter} implies that the function (\ref{defTdel}) is continuous at delay values where
\begin{equation}\label{condcase1}
\|T(j\omega,\vec\tau)\|_\infty>\interleave T_a(j\omega,\vec\tau)\interleave_\infty.
\end{equation}
This property, along with the fact that $\interleave T_a(j\omega,\vec\tau)\interleave_\infty$ is independent of $\vec\tau$ (see Proposition~\ref{prop:Tasinfprop}), lead to the assertion (\ref{shinfnorm}) and the continuity of (\ref{shinfnorm2}) under the condition (\ref{condcase1}). In the other case the assertions follow from Lemma~\ref{lem3ap}.
\hfill $\Box$

The explicit expression (\ref{shinfnorm}) lays at the basis of an algorithm to compute the strong $\Hi$ norm presented in the accompanying paper \cite{TW579:10}.
\smallskip

\section{Numerical Example} \label{sec:ex}

\begin{figure}[!h]
    \begin{minipage}[t]{0.45\textwidth}
        \vspace{0pt}\raggedright
        \begin{center}
        \includegraphics[width=0.7\linewidth]{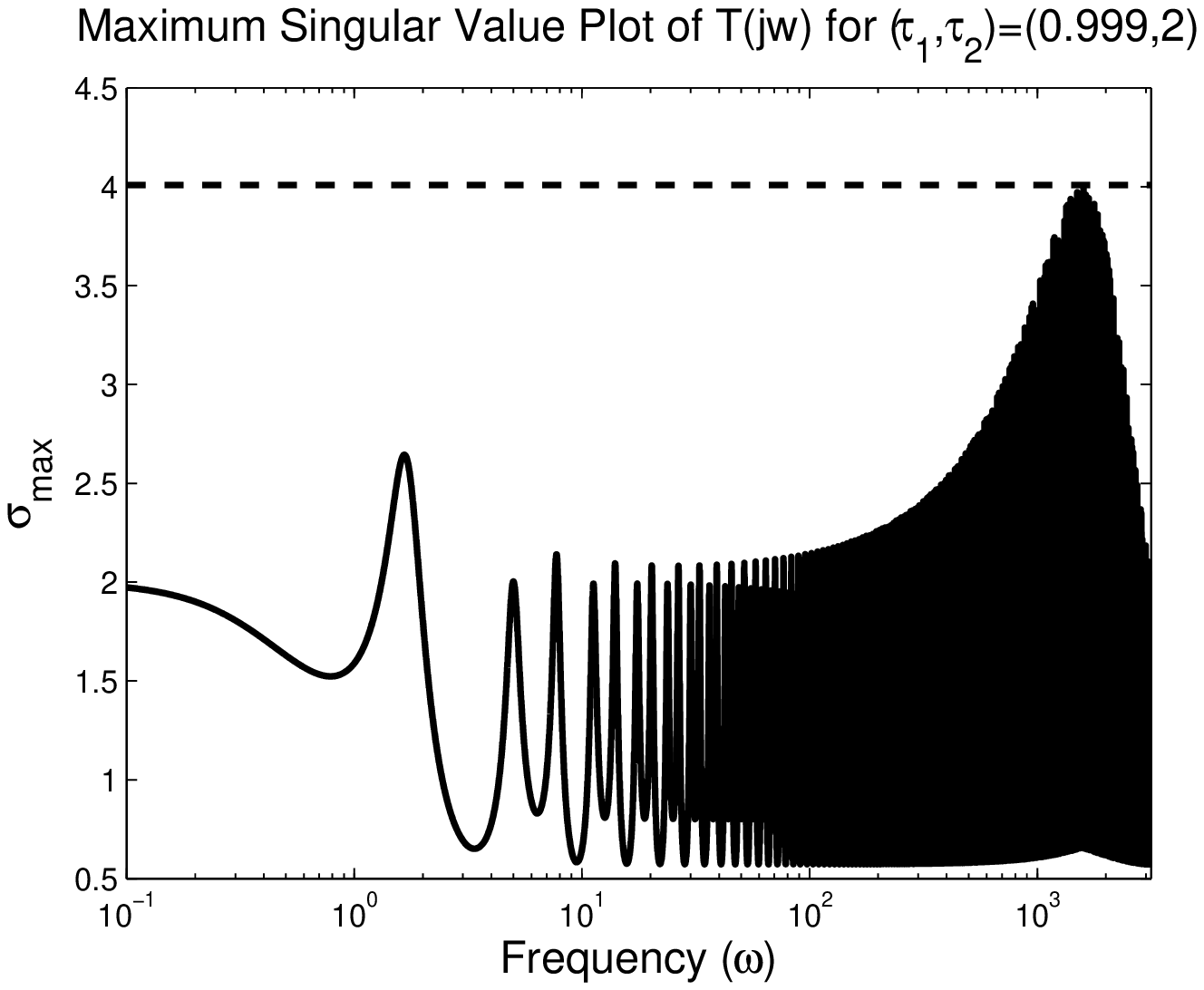}
        \caption{\label{fig:hinfalg_nox} The maximum singular value plot of $T$ (\ref{Tex}): $\|T(j\w,\vec\tau)\|_{\infty}<\interleave T_a(j\w,\vec \tau)\interleave_\infty$ case.}
        \end{center}
   \end{minipage}
\hfill
    \begin{minipage}[t]{0.45\textwidth}
        \vspace{0pt}
        \begin{center}
        \includegraphics[width=0.7\linewidth]{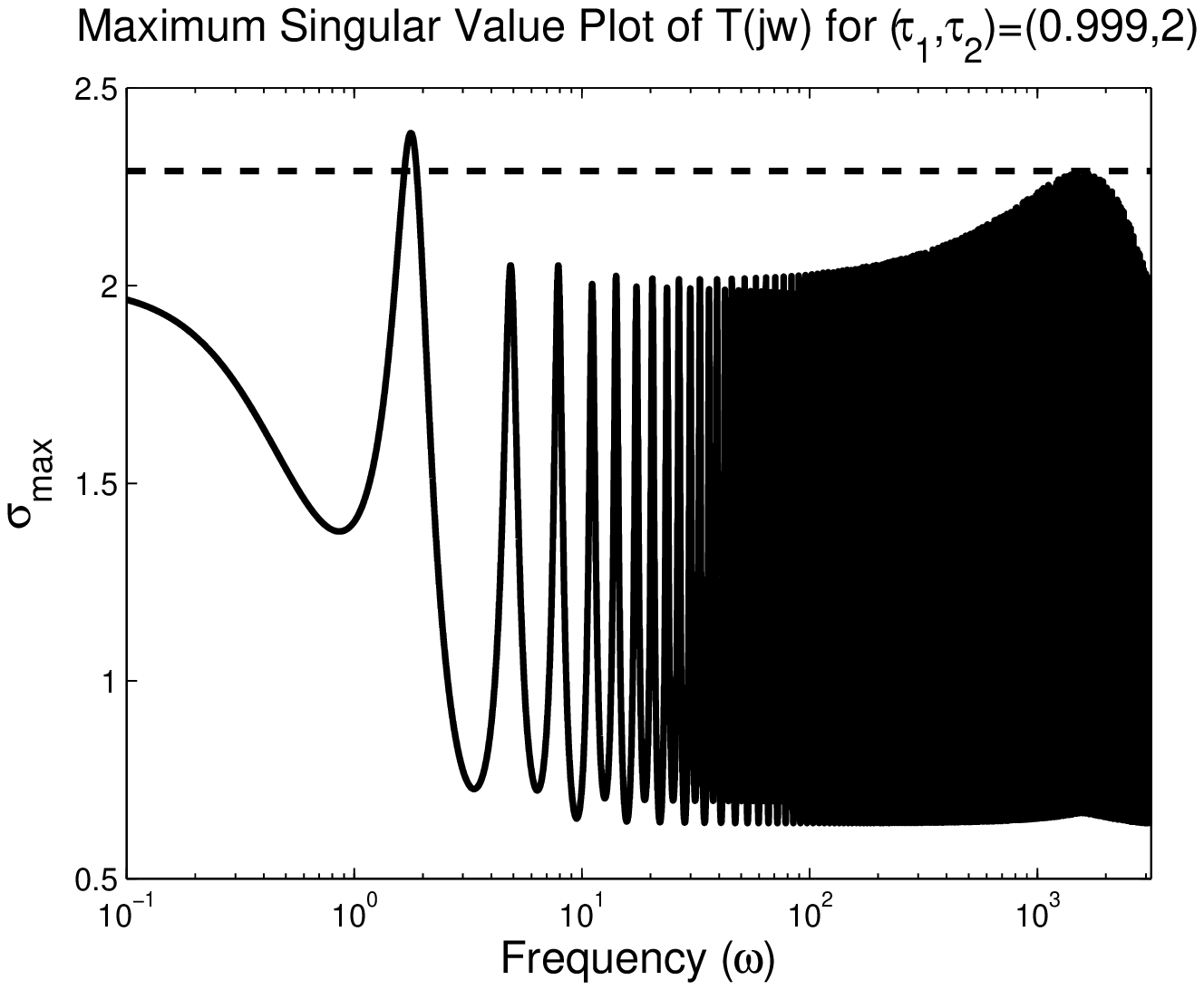}
        \caption{\label{fig:hinfalg_wx} The maximum singular value plot of $T$ (\ref{Tex2}): $\|T(j\w,\vec\tau)\|_{\infty}>\interleave T_a(j\w,\vec \tau)\interleave_\infty$ case.}
        \end{center}
        \end{minipage}
\end{figure}

By (\ref{shinfnorm}), the strong $\Hi$ norm of the transfer function $T$ is determined by either the $\Hi$ norm of $T$ or the strong $\Hi$ norm of $T_a$. We illustrate both cases.

Given the transfer function $T$ (\ref{Tex}), the strong $\Hi$ norm of its asymptotic transfer function $T_a$ is equal to $4$ (indicated as a dashed line) and the $\Hi$ norm of $T$ is $2.6422$ as shown in Figure~\ref{fig:hinfalg_nox}. Then the strong $\Hi$ norm of $T$ (\ref{Tex}) is equal to the strong $\Hi$ norm of (\ref{Taex}), namely $4$.

As a second example, consider the transfer function
\begin{equation} \label{Tex2}
T(\lambda,\vec\tau):=\frac{\lambda+2}{\lambda(1-1/16e^{-\lambda\tau_1}+1/2e^{-\lambda\tau_2})+1},
\end{equation}
with $\vec\tau=(1,2)$, and its asymptotic transfer function
\begin{equation} \label{Taex2}
T_a(\lambda,\vec\tau):=\frac{1}{(1-1/16^{-\lambda\tau_1}+1/2e^{-\lambda\tau_2})}.
\end{equation}
Figure~\ref{fig:hinfalg_wx} shows that the strong $\Hi$ norm of $T$ (\ref{Tex2}) is equal to the $\Hi$ norm of $T$ (\ref{Tex2}). Note that the strong $\Hi$ norm of the asymptotic transfer function can be used as the first level to compute the strong $\Hi$ norm in well-known level set methods \cite{boydbala,steinbuch}.

\section{Concluding Remarks} \label{sec:conc}
We analyzed the sensitivity of the $\Hi$ norm of interconnected systems with time-delays. We showed that a very broad class of interconnected  retarded and/or neutral systems can be brought in the standard form (\ref{system}) in a systematic way. Input/output delays and direct feedthrough terms can be dealt with by introducing slack variables. An additional advantage in the context of control design is the linearity of the closed loop matrices w.r.t. the controller parameters.

We showed the sensitivity of the $\Hi$ norm w.r.t.~small delay perturbations and  introduced the \emph{strong $\Hi$ norm} for DDAEs inline with the notion of strong stability. We analyzed its continuity properties derived as an explicit expression. The given properties are illustrated on numerical examples.

\section*{Acknowledgements}
This work has been supported by the Programme of Interuniversity Attraction Poles of the Belgian Federal Science Policy Office (IAP P6- DYSCO), by OPTEC, the Optimization in Engineering Center of the K.U.Leuven, by the project STRT1-09/33 of the K.U.Leuven Research Council and the project G.0712.11N of
the Research Foundation - Flanders (FWO).

\bibliography{referentielijst,otherref}            % bib file to produce the bibliography

\begin{thebibliography}{4}
\providecommand{\natexlab}[1]{#1}
\providecommand{\url}[1]{\texttt{#1}}
\providecommand{\urlprefix}{URL }
\expandafter\ifx\csname urlstyle\endcsname\relax
  \providecommand{\doi}[1]{doi:\discretionary{}{}{}#1}\else
  \providecommand{\doi}{doi:\discretionary{}{}{}\begingroup
  \urlstyle{rm}\Url}\fi

\bibitem[{Able(1956)}]{Abl:56}
Able, B. (1956).
\newblock Nucleic acid content of microscope.
\newblock \emph{Nature}, 135, 7--9.

\bibitem[{Able et~al.(1954)Able, Tagg, and Rush}]{AbTaRu:54}
Able, B., Tagg, R., and Rush, M. (1954).
\newblock Enzyme-catalyzed cellular transanimations.
\newblock In A.~Round (ed.), \emph{Advances in Enzymology}, volume~2, 125--247.
  Academic Press, New York, 3rd edition.

\bibitem[{Keohane(1958)}]{Keo:58}
Keohane, R. (1958).
\newblock \emph{Power and Interdependence: World Politics in Transitions}.
\newblock Little, Brown \& Co., Boston.

\bibitem[{Powers(1985)}]{Pow:85}
Powers, T. (1985).
\newblock Is there a way out?
\newblock \emph{Harpers}, 35--47.

\end{thebibliography}


\begin{thebibliography}{13}
\providecommand{\natexlab}[1]{#1}
\providecommand{\url}[1]{\texttt{#1}}
\providecommand{\urlprefix}{URL }
\expandafter\ifx\csname urlstyle\endcsname\relax
  \providecommand{\doi}[1]{doi:\discretionary{}{}{}#1}\else
  \providecommand{\doi}{doi:\discretionary{}{}{}\begingroup
  \urlstyle{rm}\Url}\fi

\bibitem[{Boyd and Balakrishnan(1990)}]{boydbala}
Boyd, S. and Balakrishnan, V. (1990).
\newblock A regularity result for the singular values of a transfer matrix and
  a quadratically convergent algorithm for computing its
  $\mathcal{L}_{\infty}$-norm.
\newblock \emph{Systems \& Control Letters}, 15, 1--7.

\bibitem[{Bruinsma and Steinbuch(1990)}]{steinbuch}
Bruinsma, N. and Steinbuch, M. (1990).
\newblock A fast algorithm to compute the $\mathcal{H}_{\infty}$-norm of a
  transfer function matrix.
\newblock \emph{Systems and Control Letters}, 14, 287--293.

\bibitem[{Burke et~al.(2006)Burke, Henrion, Lewis, and Overton}]{Burke-hifoo}
Burke, J.V., Henrion, D., Lewis, A.S., and Overton, M.L. (2006).
\newblock {HIFOO} - a {\sc matlab} package for fixed-order controller design
  and {H}-infinity optimization.
\newblock In \emph{Proceedings of the 5th IFAC Symposium on Robust Control
  Design}. Toulouse, France.

\bibitem[{Fridman and Shaked(2002)}]{fridman}
Fridman, E. and Shaked, U. (2002).
\newblock {$H_{\infty}$}-control of linear state-delay descriptor systems: an
  lmi approach.
\newblock \emph{Linear Algebra and its Applications}, 351-352, 271--302.

\bibitem[{Gumussoy and Michiels(2010)}]{TW579:10}
Gumussoy, S. and Michiels, W. (2010).
\newblock Fixed-order strong h-infinity control of interconnected systems with
  time-delays.
\newblock \emph{submitted to $18^\textrm{th}$ World Congress of IFAC 2011. See
  also Technical Report TW579, Department of Computer Science, K.U.Leuven,
  2010.}

\bibitem[{Hale and Verduyn~Lunel(2002)}]{have:02}
Hale, J. and Verduyn~Lunel, S. (2002).
\newblock Strong stabilization of neutral functional differential equations.
\newblock \emph{IMA Journal of Mathematical Control and Information}, 19,
  5--23.

\bibitem[{Michiels et~al.(2002)Michiels, Engelborghs, Roose, and
  Dochain}]{TW-report-286}
Michiels, W., Engelborghs, K., Roose, D., and Dochain, D. (2002).
\newblock Sensitivity to infinitesimal delays in neutral equations.
\newblock \emph{{SIAM} Journal on Control and Optimization}, 40(4), 1134--1158.

\bibitem[{Michiels and Gumussoy(2010)}]{wimsimax}
Michiels, W. and Gumussoy, S. (2010).
\newblock Characterization and computation of h-infinity norms of time-delay
  systems.
\newblock \emph{{SIAM} Journal on Matrix Analysis and Applications}, 31(4),
  2093--2115.

\bibitem[{Michiels and Niculescu(2007)}]{bookwim}
Michiels, W. and Niculescu, S.I. (2007).
\newblock \emph{Stability and stabilization of time-delay systems. An
  eigenvalue based approach}.
\newblock SIAM.

\bibitem[{Michiels and Vyhl{\'\i}dal(2005)}]{Michiels:2005:NEUTRAL}
Michiels, W. and Vyhl{\'\i}dal, T. (2005).
\newblock An eigenvalue based approach for the stabilization of linear
  time-delay systems of neutral type.
\newblock \emph{Automatica}, 41(6), 991--998.

\bibitem[{Michiels et~al.(2009)Michiels, Vyhl{\'\i}dal, Z{\'\i}tek, Nijmeijer,
  and Henrion}]{Michiels:2007:MULTIVARIATE}
Michiels, W., Vyhl{\'\i}dal, T., Z{\'\i}tek, P., Nijmeijer, H., and Henrion, D.
  (2009).
\newblock Strong stability of neutral equations with an arbitrary delay
  dependency structure.
\newblock \emph{{SIAM} Journal on Control and Optimization}, 48(2), 763--786.

\bibitem[{Stykel(2002)}]{stykel}
Stykel, T. (2002).
\newblock On criteria for asymptotic stability of differential algebraic
  equations.
\newblock \emph{{ZAMM} Z. Angew. Math. Mech.}, 82(3), 147--158.

\bibitem[{Zhou et~al.(1995)Zhou, Doyle, and Glover}]{zhou}
Zhou, K., Doyle, J., and Glover, K. (1995).
\newblock \emph{Robust and optimal control}.
\newblock Prentice Hall.

\end{thebibliography}

\section{Some technical lemmas}\label{sec:Appendix2}

\begin{lem}\label{propconverge2}
For all $\gamma>0$, there exist numbers $\epsilon>0$ and  $\Omega>0$ such that
\[
\sigma_{1}\left(T(j\w,\vec r)-T_a(j\w,\vec r)\right)<\gamma
\]
 fir all $\w>\Omega$ and $\vec r\in\mathcal{B}(\vec\tau,\epsilon)\cap(\mathbb{R}^+)^m$.
\end{lem}
\noindent\textbf{Proof.\ }
   The uniformity of the bound $\gamma$ w.r.t.~small delay perturbations  stems from the fact that the bound (\ref{normA22}) is a continuous function of the delays $\vec \tau$ at their nominal values. The latter is implied  by  the \emph{strong} stability assumption (Assumption~\ref{assumption_sstab}). \hfill $\Box$

\begin{lem} \label{lem:fininter}
Let $\xi>\interleave T_a(j\w,\vec \tau)\interleave_\infty$ hold.
Then there exist real numbers $\epsilon>0,\ \Omega>0$ and an integer $N$ such that for any $\vec r\in\mathcal{B}(\vec\tau,\epsilon)\cap(\mathbb{R}^+)^m$, the number of frequencies $\w^{(i)}$ such that
\begin{equation}
\sigma_k\left(T(j\w^{(i)},\vec r)\right)=\xi,
\end{equation}
for some $k\in\{1,\ldots,n\}$, is smaller then $N$, and, moreover, $|\w^{(i)}|<\Omega$.
\end{lem}

\noindent\textbf{Proof.\ }
 For any (fixed) value of $\xi>0$ and delays $\vec r$,  the relation
 \begin{equation}\label{omhold}
 \sigma_k\left(T(j\w),\vec r\right)=\xi
\end{equation}
 holds for some $\omega\in\mathbb{R}$ and $k\in\{1,\ldots,n\}$ if and only if $\lambda=j\omega$ is a zero of the function
{\small \begin{equation}\label{ham}
\det\left(\left[\begin{array}{cc}
\lambda E-A_0-\sum_{i=1}^m A_i e^{-\lambda r_i} & -\frac{1}{\xi}B B^T \\
\frac{1}{\xi}CC^T & \lambda E^T+A_0^T +\sum_{i=1}^m A_i^T e^{\lambda r_i}
\end{array}\right]\right).
\end{equation}}
This result is a variant of Lemma~2.1 of \cite{wimsimax} to which we refer for the proof.

Now take $\xi>\interleave T_a(j\w,\vec \tau)\interleave_\infty$. From Lemma~\ref{propconverge2}, and taking into account that $\interleave T_a(j\w,\vec \tau)\interleave_\infty$ does not depend on $\vec\tau$ (see Proposition~\ref{prop:Tasinfprop}) it follows that there exists numbers  $\epsilon>0$ and $\Omega>0$ such that all $\omega$ satisfying (\ref{omhold}) for some $k\in\{1,\ldots,n\}$ and $\vec r\in\mathcal{B}(\vec\tau,\epsilon)\cap(\mathbb{R}^+)^m$ also satisfy
$
|\omega|< \Omega.
$
This proves one statement. At the same time $\lambda=j\omega$ must be a zero of the analytic function (\ref{ham}). The other statement is due to the fact that an analytic function only has finitely many zeros in a compact set. \hfill $\Box$

\begin{lem}\label{lem3ap}
The following implication holds \\
$\| T(j\w,\vec \tau)\|_{\infty} \leq  \interleave T_a(j\w,\vec \tau)\interleave_\infty \ \Rightarrow$ \\
${}\hspace{3.9cm}\interleave T(j\w,\vec \tau)\interleave_{\infty} = \interleave T_a(j\w,\vec \tau)\interleave_\infty.
$
\end{lem}

\noindent\textbf{Proof.\ }
For every $\epsilon>0$ there exist delays $\vec\tau_0$ and a frequency $\omega_0$ such that
\[
\|\vec\tau_0-\vec\tau\|<\epsilon/2,\ \ \ \sigma_{1}\left(T_a(j\omega_0,\vec\tau_0)\right)\geq \interleave T_a(j\w,\vec \tau)\interleave_\infty-\epsilon/2.
\]
In addition, there exist commensurate delays
\begin{equation}\label{comdelays}
\vec\tau_r=\left(n_1/s,\ldots,n_m/s\right),
\end{equation}
with $(n_1,\ldots,n_m,s)\in\mathbb{N}^{m+1}$ such that
\begin{multline*}
\|\vec\tau_r-\vec\tau_0\|<\epsilon/2, \\ 
\ \ \ \left|\sigma_{1}\left(T_a(j\omega_0,\vec\tau_r)\right)-
\sigma_{1}\left(T_a(j\omega_0,\vec\tau_0)\right)\right|\leq\epsilon/2.
\end{multline*}
Thus, for all $\epsilon>0$ there exist commensurate delays (\ref{comdelays}) and a frequency $\omega_0$ satisfying
\[
 \|\vec\tau_r-\vec\tau\|<\epsilon,\ \
\sigma_{1}\left(T_a(j\omega_0,\vec\tau_r)\right)\geq \interleave T_a(j\w,\vec \tau)\interleave_\infty-\epsilon.
\]
From the fact that
\[
T_a(j\omega_0,\vec\tau_r)=T_a\left(j(\omega_0+2\pi s k) ,\vec\tau_r\right)
\]
for all $k\geq 1$ and Lemma~\ref{propconverge2}, we conclude that
\begin{equation}\label{leftright}
\interleave T(j\w,\vec \tau)\interleave_{\infty} \geq  \interleave T_a(j\w,\vec \tau)\interleave_\infty.
\end{equation}

Now take a level $\xi> \interleave T_a(j\w,\vec \tau)\interleave_\infty$, and let $\epsilon$ and $\Omega$ be determined by the assertion of Lemma~\ref{lem:fininter}. From the assumption $\| T(j\w,\vec \tau)\|_{\infty} \leq  \interleave T_a(j\w,\vec \tau)\interleave_\infty$  and the relation between (\ref{omhold}) and (\ref{ham}) it follows that the function (\ref{ham}) has no zeros on the imaginary axis for $\vec r=\vec\tau$. Because the function  (\ref{ham}) is analytic and all potential imaginary axis zeros have modulus smaller than $\Omega$ whenever $\vec r\in\mathcal{B}(\vec\tau,\epsilon)\cap(\mathbb{R}^+)^m$, we conclude that there exists a number $\epsilon_2>0$ such that the function (\ref{ham}) has no imaginary axis eigenvalues whenever $\vec r\in\mathcal{B}(\vec\tau,\epsilon_2)\cap(\mathbb{R}^+)^m$. Equivalently, $T(j\omega,\vec r)$ has no singular values equal to $\xi$ whenever $\vec r\in\mathcal{B}(\vec\tau,\epsilon_2)\cap(\mathbb{R}^+)^m$.  This proves that the  left and the right hand side of (\ref{leftright}) are equal. \hfill $\Box$
\end{document}